\documentclass{aastex}
\usepackage{spr-astr-addons}
\usepackage{url}

\RequirePackage{color}

\newcommand{\emaila}{zhangyong5@mail.sysu.edu.cn}

\begin{document}

\title{Hydrogenated Fullerenes (Fulleranes) in Space}
\shorttitle{Fulleranes}
\shortauthors{Zhang et al.}

\author{Yong Zhang\altaffilmark{1,2}}
\email{\emaila}
\and \author{Seyedabdolreza Sadjadi\altaffilmark{1,2}}
\and \author{Chih-Hao Hsia\altaffilmark{3}}

\altaffiltext{1}{School of Physics and Astronomy, Sun Yat-Sen University, Zhuhai, China}
\altaffiltext{2}{Laboratory for Space Research, The University of Hong Kong, Hong Kong, China}
\altaffiltext{3}{State Key Laboratory of Lunar and Planetary Sciences, Macau University of Science and Technology, Macau, China}

\begin{abstract}

Since the first laboratory synthesis of C$_{60}$ in 1985, fullerene-related species
have been proposed to interpret various astronomical features. After more than 25 years' efforts, 
several circumstellar and interstellar features have been convincingly assigned to
C$_{60}$, C$_{70}$, and C$_{60}^+$.  These successes resulted from the recent advancements
in observational, experimental, as well as computational techniques, and  re-stimulated
interest in searching for fullerene derivatives in space. As one of the most important fullerene derivatives,
hydrogenated fullerene (fullerane) is likely to exist in circumstellar and interstellar conditions.
This review gives an overview of the chemical properties and spectral signals of fulleranes focusing  on
those relevant to astronomy. We summarize previous proposals of fulleranes as the carrier 
of astronomical features at UV, optical, infrared, and radio wavelengths, and discuss the arguments 
favoring or disfavoring  the presence of fulleranes in  astronomical environments.
Although no unambiguous detection of fulleranes in space has yet been reported, there are plausible
evidences for supporting the formation of certain fullerane isomers.

\end{abstract}



\section{Introduction}

The unidentified infrared emission (UIE) bands have been detected in circumstellar envelopes and 
the interstellar medium (ISM). Although most of the UIE bands are attributed to  vibrational modes of 
$sp^2$ and  $sp^3$ C--H and C--C bonds, their exact carrier has long been an enigma. Since their 
synthesis in laboratory \citep{kroto}, fullerenes have been believed to exist in significant amounts 
in space. The carbon cage structure is highly stable against intense radiation, making fullerenes
and their derivatives an attractive material candidate being partly responsible for the UIE
phenomenon.  As the dominant source of cosmic dust, stars leaving the asymptotic giant branch (AGB)
stand as the key source of searching for astronomical fullerenes. Early efforts on searching
for circumstellar fullerenes, however, were unsuccessful \citep[see the references in][]{zhang13}.

A breakthrough in the last ten years is that four UIE bands at 7.0, 8.5, 17.4, and 18.9\,$\mu$m
were properly assigned to C$_{60}$ \citep{cami10,sel10,gar10}, the most stable member of fullerene family.  
These C$_{60}$ infrared bands  were subsequently detected
in various astronomical environments \citep[see][for a review]{zhang13}.
Weaker C$_{70}$ bands were also detected in a few planetary nebulae (PNe) exhibiting C$_{60}$ 
features \citep{cami10,gar11}. The discovery of C$_{60}$ in a proto-PN suggests that 
this big molecule can be rapidly formed within a short timescale of $\sim10^3$\,yr \citep{zhang11}.
If fullerenes are able to survive the formation of solar system, they could be detected in the
pre-solar grains. However,
searches for fullerenes in meteorites have yielded conflicting results \citep[e.g.][]{be95,hey97}.
As fullerene molecules can be created under the laser conditions of studying fullerenes,
one should be cautious of false positive signals \citep{ha08}.
Another important finding in recent years is that 
the hypothesis of C$_{60}^+$ as the carrier of a few infrared diffuse interstellar bands (DIBs),
originally proposed by \citet{fo94}, was validated by the experimental spectra \citep{cam15,cam16}.
Therefore, it is becoming clear that fullerenes  can survive in harsh conditions and be widespread in space.

Although C$_{60}$, C$_{70}$, and C$_{60}^+$ are the only fullerene members that have been
convincingly detected so far in astronomical environments, the presence of other fullerene molecules 
have been extensively investigated. For instance, the possibility of buckyonions (multishell fullerenes) 
as the carrier of the interstellar extinction bump at 217.5\,nm have been discussed
by \citet{ig04} and \citet{li08}. The small fullerene C$_{24}$ was proposed to be responsible for 
the UIE band at 11.2\,$\mu$m \citep{ber17}. Through a comparison between the theoretical and observational
spectra, \citet{can19} argued that C$_{44}$, C$_{50}$, and C$_{56}$ might be present in several C$_{60}$ 
sources.  Selected elements can be trapped within the
carbon cage of fullerenes, providing a potential interpretation for the anomalous element
enrichment in stardust \citep{dunk13}. These authors found that the encapsulation of a metal within 
C$_{60}$ would cause some vibrational modes to become IR-active and produce features coincident with the 
UIE bands.

With a high electron affinities, fullerenes are chemically active. 
\citet{gar13} show that fullerenes can react with small polycyclic aromatic hydrocarbons (PAHs),
such as anthracene, to form  Diels-Alder cycloaddition products, which exhibits
spectral features similar to the UIE bands. The efficient formation of C$_{60}$/PAH
adducts was confirmed by the gas phase experiments  \citep{dunk13}.
The high stability and rich diversity of fullerene molecules make them ideal carrier candidates
of DIBs.  \citet{om16} made a thorough discussion of the likely contributions to DIBs from
various fullerene derivatives including metallofullerenes, heterofullerenes, hydrogenated
fullerenes (fulleranes), fullerene-PAH compounds, and H$_2$@C$_{60}$.

It is essentially unclear what forms of fullerene are abundant in various astronomical environments. 
As C$_{60}$ has a high reaction rate with atomic hydrogen, the most abundant element in the
universe, along with the facts that C$_{60}$-containing PNe are not H-deficient \citep{gar10},
it is logical to put its hydrogenated/protonated analogues (i.e. fulleranes) on the top of the list.
If existing, fulleranes may play a vital role on interstellar and circumstellar chemistry.
The purpose of this paper is to review past efforts made to investigate fulleranes in space.


\section{Chemistry of fulleranes}

When mixing with hydrogen atoms in solvents, C$_{60}$ can be quickly
hydrogenated into C$_{60}$H$_{36}$ \citep[see e.g.][]{cat09,igl12}. Various chemical
methods have been developed to synthesize C$_{60}$H$_{m}$ \citep[see][for a review]{bm10}.
C$_{60}$H$_{18}$ is the only species that can be produced by various methods at a high yield.  
Higher fulleranes (C$_{60}$H$_{m}$ with $m>36$) are chemically unstable, and can be synthesized only under
very severe conditions. Further hydrogenation can completely destruct C$_{60}$ cages
due to severe angle and torsional strains, and the fragments of fulleranes can
result in the formation of large PAHs. Gas-phase experiments suggested that
C$_{60}^+$ and C$_{60}^{2+}$ can react with hydrogen atoms to form C$_{60}$H$_{m}^{+}$ and 
C$_{60}$H$_{m}^{2+}$ with $m=1$--3 and these fullerene ions are unreactive with 
molecular hydrogen \citep{Petrie1992}. Larger fullerene ions, like C$_{70}^{2+}$, are
less reactive than C$_{60}^{2+}$ \citep{pb00}.

It is established that the fullerenes
obeying the isolated-pentagon rule (IPR) are more stable than adjacent-pentagon fullerenes.
C$_{60}$  stands out within the fullerene family as the smallest IPR fullerene.
\citet{pb94} found that hydrogenation can greatly reduce the chemical instability
induced by adjacent-pentagon structures. Higher energy is required to dissociate
an adjacent-pentagon C-H bond than an isolated-pentagon C-H bond. Therefore, 
hydrogen atoms tend to possess or migrate to adjacent-pentagon sites, and 
non-IPR fullerenes are more prone to be hydrogenated,
 favoring the production of smaller fulleranes with non-IPR structures in
interstellar and circumstellar environments, such as C$_{20}$H$_{20}$.

The heating treatment of C$_{60}$H$_{36}$ can cause dehydrogenation and easily release
H$_2$ molecules \citep{cat09,igl12}. In laboratory conditions, hydrogenation and dehydrogenation of 
C$_{60}$ and C$_{60}$H$_{36}$ are efficient, and the C$_{60}$ cage is highly resistant to these chemical
reactions. This leads to a hypothesis that fullerenes or fullerene-like materials may provide
a surface to catalyze the formation of molecular hydrogen in astronomical environments
\citep[see][for a review]{cat10}. Because direct gas-phase reactions are inefficient 
in the ISM, it has been recognized that catalytic reactions on surfaces of dust grains
should be responsible for the abundant interstellar H$_2$. Experiments show that colliding with dust
grains, a hydrogen atom can be readily bound to the grain surfaces through physisorption.
However, as the physisorption bond is very weak, chemisorption dominates the regions
with a dust temperature higher than 20\,K. Theoretical calculations suggest that
the addition of pentagonal rings in graphene sheets (fullerene-like structure)
can substantially reduce the activation barrier against the chemisorption of a hydrogen atom
\citep{iv10}. Once a hydrogen atom is chemisorbed, another hydrogen can  be attached  to the adjacent carbon 
atom without barrier, and then a third hydrogen hit atop the second one to form H$_2$ through the Eley-Rideal 
mechanism. Therefore, fulleranes might play an important role on the  H$_2$ formation
in warm gas. 

Through an investigation of the ion/molecule chemistry of fullerene derivatives in astronomical
environments, \citet{pb00}  showed that C$_{60}$ can form C$_{60}^{2+}$  
 through charge transfer electron detachment from He$^+$  and cosmic-ray impact.
C$_{60}^{2+}$ ions could be efficiently hydrogenated into C$_{60}$H$^{2+}$ in interstellar conditions.
Hydrogenation tends to retard the addition reactions of fullerene ions with
polar and small unsaturated molecules. Therefore, in H-rich environments, the association
of fullerene ions and interstellar molecules might be insignificant.

It is difficult to estimate the abundance ratio of C$_{60}$ and its hydrides for the reasons
outlined by \citet{om16}.  C$_{60}$H and C$_{60}$H$^+$ can be efficiently dissociated by absorbing a hard UV photon.
The models of dark clouds suggest that C$_{60}$ can be efficiently protonated so that
the fractional abundance of C$_{60}$H$^+$ is relatively larger than C$_{60}^+$ \citep{tom92}.
However, based on the formation and dissociation rates, \citet{om16} estimated that the abundance ratios
of C$_{60}$H/C$_{60}$ and C$_{60}$H$^+$/C$_{60}^+$ in the ISM should be roughly {0.1}--{0.2} 
with a large uncertainty. Given their stronger H bonding, the fulleranes with even number of H atoms 
and higher hydrogenated fullerenes are more resistant to photolysis. 
The photo-dissociation rates of these fulleranes, however, are largely unknown, precluding 
the accurate determination of the fullerane abundance.  
To  establish the environments in which fulleranes may be present,
it is instructive to compare the interstellar and circumstellar conditions 
with the dehydrogenation-model predictions of large PAHs \citep{mj13}. 
\citet{cb14} found that C$_{60}$ emission in reflection nebulae arises from the 
regions where the dust temperature ranges from 20--40\,K and large PAHs (like C$_{54}$H$_{18}$ and 
C$_{66}$H$_{20}$) are fully dehydrogenated. If C$_{60}$H$_m$ has a similar hydrogenation rate
as large PAHs, fulleranes should not be abundantly present in these regions, 
but should instead survive in the dense and UV-shielded environments (such as those of PPNe) where
PAHs are normal or super-hydrogenated \citep[see Figure~13 of][]{cb14}.

\section{Fulleranes as the carriers of interstellar and circumstellar spectral features}

\subsection{Fulleranes versus UIE bands}

Based on a force-field model of fully hydrogenated C$_{60}$, \citet{web91} suggested that slightly 
and heavily hydrogenated fullerenes may be responsible for the 3.3\,$\mu$m and 3.4\,$\mu$m UIE
bands, respectively. It is well established that the two features arise from the
$sp^2$ aromatic and $sp^3$ aliphatic C-H stretching motions \citep[see a review by][]{kwok16}.  
One may envision that with increasing hydrogen coverage on fullerene cages, the $sp^2$ carbons
should be gradually converted into $sp^3$ carbons, and thus lead to a shift of
the C-H stretching bond from 3.3\,$\mu$m to 3.4\,$\mu$m.
The calculated spectrum of C$_{60}$H$_{60}$ also exhibits a number
of peaks between 6--10\,$\mu$m associated with the rocking of C-H units, which might account 
for the 6--9\,$\mu$m plateau and the 7.7\,$\mu$m UIE band \citep{web93b}. 
A recent theoretical study showed that the inclusion of non-planar structural defects in aromatic core molecular structures may account for the UIE
bands, especially those lying in the 6--9\,$\mu$m range \citep{gl17}.
Fullerane fragments have a similar molecular structure, and thus are a promising
UIE carrier.

An experimental study shows that when exposed to atomic hydrogen, C$_{60}$ can produce
a number of mid-infrared features resembling the UIE bands \citep{sto01}. Specifically, a band
at 7.63\,$\mu$m grows with increasing hydrogen exposure, but its wavelength does not shift,
which is compatible with the conjecture of fulleranes as the carrier of the 7.7\,$\mu$m feature
\citep{web93b}. Moreover, \citet{cat03} found that the absorption spectrum of C$_{60}$H$_{36}$ can 
match several UIE bands detected in  proto-PNe.
The experimental spectra also reveal a C-H stretching band peaking at 3.33--3.45\,$\mu$m 
that is reasonably consistent with astronomical observations of the 3.4\,$\mu$m feature \citep{sto01}. 
However, the position of the C-H stretching band does not shift to 
3.3\,$\mu$m even for slightly hydrogenated C$_{60}$,  in contrast to the proposal of 
\citet{web91}.  Therefore, the 3.3\,$\mu$m aromatic feature is unlikely to be carried by fulleranes.
Strictly speaking, fullerenes cannot be regarded as pure aromatic molecules. The pentagon related 
curvature induces a slight admixture of $sp^3$ character.

\citet{web95}  proposed that C$_{60}$H$_m$ might be responsible for the so-called 21\,$\mu$m feature,
a rare UIE band only observed in a few carbon-rich proto-PNe. 
The 21\,$\mu$m feature was first discovered with {\it IRAS} in four proto-PNe \citep{kv89}, and now has 
been detected in 27 objects \citep{ml16}. The {\it ISO} observations showed that this feature
actually peaks at 20.1\,$\mu$m, and its profile is remarkably consistent among different sources
\citep{vol99}. However, recent {\it Spitzer}/{\it IRS} spectra suggested a central wavelength of
$20.47\pm0.10$\,$\mu$m \citep{sloan}. Intriguingly, a strong feature peaking at 21\,$\mu$m was 
recently revealed in two supernova remnants although its observational properties differ from those
of the 21\,$\mu$m feature in proto-PNe \citep{rho18}.  It is noticeable that the experimental and computational  spectra of fulleranes exhibit prominent features in the 18--22\,$\mu$m spectral range
\citep{igl12,zhang17}. Theoretical calculations show that
with increasing hydrogen coverage, the central wavelength monotonically shifts from 18\,$\mu$m to 22\,$\mu$m
\citep{web95,zhang20}. A weighted combination of the theoretical spectra of various fullerane isomers
can, in principle,  reproduce the observed 21\,$\mu$m feature.
If the 21\,$\mu$m feature could be attributed to C$_{60}$H$_m$, \citet{zhang20}
found that the degree of hydrogenation must be intermediate ($m=$10--20). 
The main issue of the fullerane hypothesis is to interpret 
the unvaried shape of the 21\,$\mu$m feature.  Probably, only stable fulleranes having
`magic numbers' of hydrogen atoms favored by symmetry considerations can survive in the short
evolutionary timescale, as suggested by \citet{web95}. 
The hypothesis of PAH molecules as the carrier of other
UIE bands faces the same problem  \citep{kz13}, but even more seriously, since the other UIE bands, unlike
the 21\,$\mu$m feature, have been detected in more diverse astronomical environments.

\subsection{Fulleranes versus extended red emission}

Extended red emission (ERE) is a broad featureless emission band spanning  the wavelength range from
about 5000--9000\,{\AA} \citep[see a review by][]{wv04}. Since the first discovery in Red Rectangle nebula
\citep{co75}, ERE has been detected in a variety of extended objects, including
reflection nebulae, PNe, \ion{H}{2} regions, and the high-latitude diffuse ISM. It has a full-width-half-maximum
of 600--1200\,{\AA} with a long red emission tail.  The peak wavelength varies from 6000--8500\,{\AA}
in different regions. Although the carrier of ERE remains unidentified, a general consensus is that
ERE arises from carbonaceous materials, including
hydrogenated amorphous carbon, quenched carbonaceous composite, nanodiamonds, etc.,
 through a photoluminescence process \citep[see][and references therein]{la17}.  
The most recently proposed ERE carrier is graphene oxide nanoparticles \citep{sar19}.
Based on a measurement of ERE in the diffuse ISM, \citet{go98} estimated a strict lower limit of the 
ERE photon conversion efficiency of $10\%\pm3\%$.  

\citet{web93c} noted the ERE spectrum of reflection nebulae resembles the laboratory spectrum of 
C$_{60}$ photoluminescence, and thus suggested that C$_{60}$, C$_{60}$H$_m$, as well as 
their ions might be the carrier of ERE. However, this proposal was ruled out by 
\citet{wv04} because  C$_{60}$ and C$_{60}^+$ were not discovered in space at that time
and the quantum efficiency of  C$_{60}$ is orders of magnitude lower than that required 
to produce ERE. Now the situation has changed with definite detection of  C$_{60}$ and
 C$_{60}^+$ in circumstellar and interstellar environments. Perhaps it is possible that the photoluminescence
efficiency could be increased through  hydrogenating C$_{60}$ to certain C$_{60}$H$_m$ isomers.
It would be attractive to examine the photoluminescence spectra of
fulleranes in laboratory, which are very scarce.

\subsection{Fulleranes versus anomalous microwave emission}

Anomalous microwave emission (AME) is a continuum emission excess above synchrotron, 
free-free, cosmic microwave background, and thermal dust in the frequency range
of 10--60\,GHz \citep{ko96,le97}.  It peaks at about 30\,GHz and has been detected
in \ion{H}{2} regions, molecular and dust clouds, supernova remnants, external galaxies
\citep[see][for a recent review]{dic18}, and recently in proto-planetary disks \citep{gre18},
suggesting that AME holds vital clues for understanding the material cycle between
ISM, stars, and planets. However, the exact mechanism responsible for AME is not yet known.
A widely accepted hypothesis is that AME arises from electric dipole radiation from rapidly-rotating 
nanoparticles \citep{dl98}, but the nature of these particles was not specified.  
PAHs can be ruled out as the AME carrier because  PAHs presumably emit strong bands at 8 and
12\,$\mu$m which, however, are not observationally associated with the AME \citep{hd16}.

The C-H bonds on the surface of fulleranes can induce a net dipole moment, and thus produce
electric dipole radiation. \citet{igl05} calculated the rotation spectra of a variety of
fulleranes and found that the emission peaks in the 1--65\,GHz range and shifts toward
lower frequencies with increasing molecular size and increasing hydrogenation degree.
Therefore, a mixture of fulleranes with appropriate size distribution and hydrogenation 
degree is able to reproduce the AME. Through a case study of a dark cloud, \citet{igl06} found that to 
explain the AME observations the fulleranes should contain 60--80 carbon atoms and have a hydrogenation
degree of ${\rm C}:{\rm H}\approx3:1$. The proto-planetary disks with AME detection also host
hydrogenated nanodiamonds \citep{gre18}.  Fulleranes and nanodiamonds
have a similar hybridization structure, and might share a similar formation processes.
 To elucidate the origin of AME, it is important to 
investigate the correlation between the infrared bands and the AME strengths, which
would be interesting projects for future {\it James Webb Space Telescope (JWST)} observations.

\subsection{Fulleranes versus the 217.5\,nm extinction bump}

The extinction curve of the Galaxy is characterized by a conspicuous and ubiquitous bump centered
at 217.5\,nm. In recent years, there has been increasing interests in investigating 
the 217.5\,nm extinction bump in high-redshift objects \citep[e.g.][]{hei19}.
Although the bump was discovered more than 50 years ago \citep{ste65}, its nature
is still not properly understood. A generally accepted view is that it 
stems from  the $\pi$--$\pi^*$ transition of carbonaceous materials. Despite exhibiting
strong UV absorption, C$_{60}$ can be ruled out as the carrier of the 217.5\,nm bump 
in that some sub-features produced by C$_{60}$ are not seen in the extinction curve \citep{kr90}.
\citet{web93a,web97} found that the spectra of C$_{60}$H$_2$ and C$_{60}$H$_4$ exhibit
stronger UV absorption and weaker sub-features compared to that of C$_{60}$, and thus
proposed that some fullerane family might be responsible for the 217.5\,nm bump.
The experimental spectra of C$_{60}$H$_{36}$ and its deuterated analogous show that
the UV absorption cross-sections of fulleranes are significantly higher than that of C$_{60}$,
and can match the 217.5\,nm bump in wavelength, width, and shape \citep{cat09}.
It was shown that with dehydrogenation of fulleranes the absorption maximum will
shift from  217.5\,nm to longer wavelength and the absorption coefficient will
gradually decrease.  Therefore, moderately hydrogenated C$_{60}$, if present in the
ISM, may substantially contribute to the 217.5\,nm bump. The same mixture
of fulleranes has been taken to interpret the AME \citep{igl06}.

\subsection{Fulleranes versus DIBs}

DIBs refer to a class of absorption features detected in in our Galaxy and other galaxies
at optical and near infrared wavelengths \citep[see][for a recent review]{geb16}. 
Since the first detection by \citet{heg22}, the origin of DIBs has been a century old puzzle.
A variety of species have been proposed as the carrier of DIBs, including fulleranes \citep{web92a}
and C$_{60}^+$ \citep{fo94}. The successful assignment of a few DIBs to  C$_{60}^+$
\citep{cam15,cam16} has rekindled the interests of investigating fullerene derivatives as the
DIB carrier \citep{om16}. With a great physical stability and a high chemical activity,
fullerenes and their analogues \citep[see e.g.][]{ig07} could reproduce the observed
pattern of DIBs.
The number of hydrogen atoms may vary depending on different interstellar environments,
and thus produce different DIBs.
Although there is no complete coincidence between fullerane features and DIBs,
the spectra of C$_{60}$H$_2$ and C$_{60}$H$_4$ exhibit a strong peak
near 4350\,{\AA}, which is comparable to a broad extinction feature centred near 4300--4400\,{\AA} 
\citep{web97}. The main challenge in this area is the vast number of fullerane isomers with only a few having
laboratory spectra. 
The DIB strengths are not correlated with those of the 217.5\,nm bump
\citep{xiang11}. If attributing both to fulleranes, the fulleranes producing
DIBs and the 217.5\,nm bump should be in different phase and/or have different hydrogenation 
degrees.

\section{Searching for fulleranes in astronomical environments}

So far, all attempts of searching for astronomical fulleranes have not yielded unambiguous results.
\citet{be95} and \citet{bb97} reported a positive detection of fulleranes in a sample of 
the Allende meteorite, which however was questioned by \citet{hey97} who
did not discover extractable C$_{60}$H$_2$ in Allende. It remains unclear why different
research groups found different results. A possible explanation is that fulleranes
are heterogeneously distributed throughout the Allende meteorite in a trace level.

 \citet{web92b}  suggested that the C-H 3.4\,$\mu$m feature could shift toward longer wavelengths
due to the effect of the vacancy of the neighboring C-H bonds. As each C-H bond of fulleranes has 
three neighboring carbon atoms, there are four possible configurations, namely 0, 1, 2, and 3
neighboring carbon bonded with hydrogen. It follows that four peaks are expected to 
appear at the wavelength range from 3.4--3.6\,$\mu$m, and their relative strengths can reflect
the degree of hydrogenation.  Indeed, these peaks can be seen in the experimental
spectra of C$_{60}$H$_{18}$  and C$_{60}$H$_{36}$
with a integrated molar absorptivity  comparable to  those of C$_{60}$ mid-infrared bands \citep{igl12}. Therefore, if
these fulleranes have a similar abundance as  C$_{60}$, they should be detectable through the
C-H stretching features.

\begin{figure}[tb]
\includegraphics[width=\columnwidth]{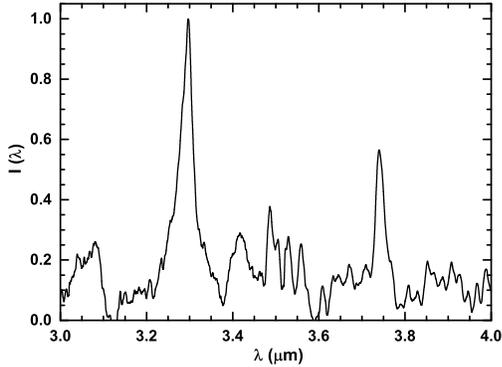}
\caption{%
The stacked {\it ISO}/{\it SWS} spectrum of six C$_{60}$ sources
(HR\,4049, IC\,418, IRAS\,01005+7910, NGC\,7023, XX Oph, and V854 Cen) in the 3--4\,$\mu$m range
} 
\label{3um}
\end{figure}

Six C$_{60}$-containing sources have {\it ISO}/{\it SWS} spectra.  Using signal-to-noise ratios
as weights, we stacked the 3--4\,$\mu$m spectra of the six sources, as shown in Figure~\ref{3um}.
A few weak features are visible in the 3.4--3.6\,$\mu$m, although we cannot uniquely attribute 
them to C$_{60}$H$_m$.
\citet{zhang13} marginally detected four peaks at 3.40, 3.48, 3.51, and 3.58\,$\mu$m
in the spectrum of a C$_{60}$ proto-PN. If the four peaks originate from C$_{60}$H$_m$,
their relative intensities indicate a moderate hydrogenation with $m=20$--40. This 
is consistent with the experimental results that 
C$_{60}$H$_{36}$ are produced at high yield and can be further heated to form
C$_{60}$H$_{18}$ \citep[see e.g.][]{cat09,igl12}.

However, the C$_{60}$-containing PNe do not exhibit prominent C-H stretching bands at 
3.4-3.6\,$\mu$m, in contrast to the expectation that fulleranes  should coexist with C$_{60}$ in 
hydrogen-rich environments. \citet{dia16} carried out a deep  spectroscopic study 
of two Galactic PNe exhibiting strong C$_{60}$ features, but failed to detect any feature
in the 3.4-3.6\,$\mu$m spectral range. Since hydrogenation of C$_{60}$  has been proven to be reversible  
\citep{cat09,igl12}, these observations suggest that the C-H aliphatic bonds can be quickly destroyed by 
shocks and/or UV radiation during the PN stage. Therefore, if existing, fulleranes
in PNe presumably have a relatively low hydrogenation degree \citep[see also][]{dia18}, and thus
are not able to produce intense bands at 3.4\,$\mu$m.   

\begin{figure*}[tb]
\includegraphics[]{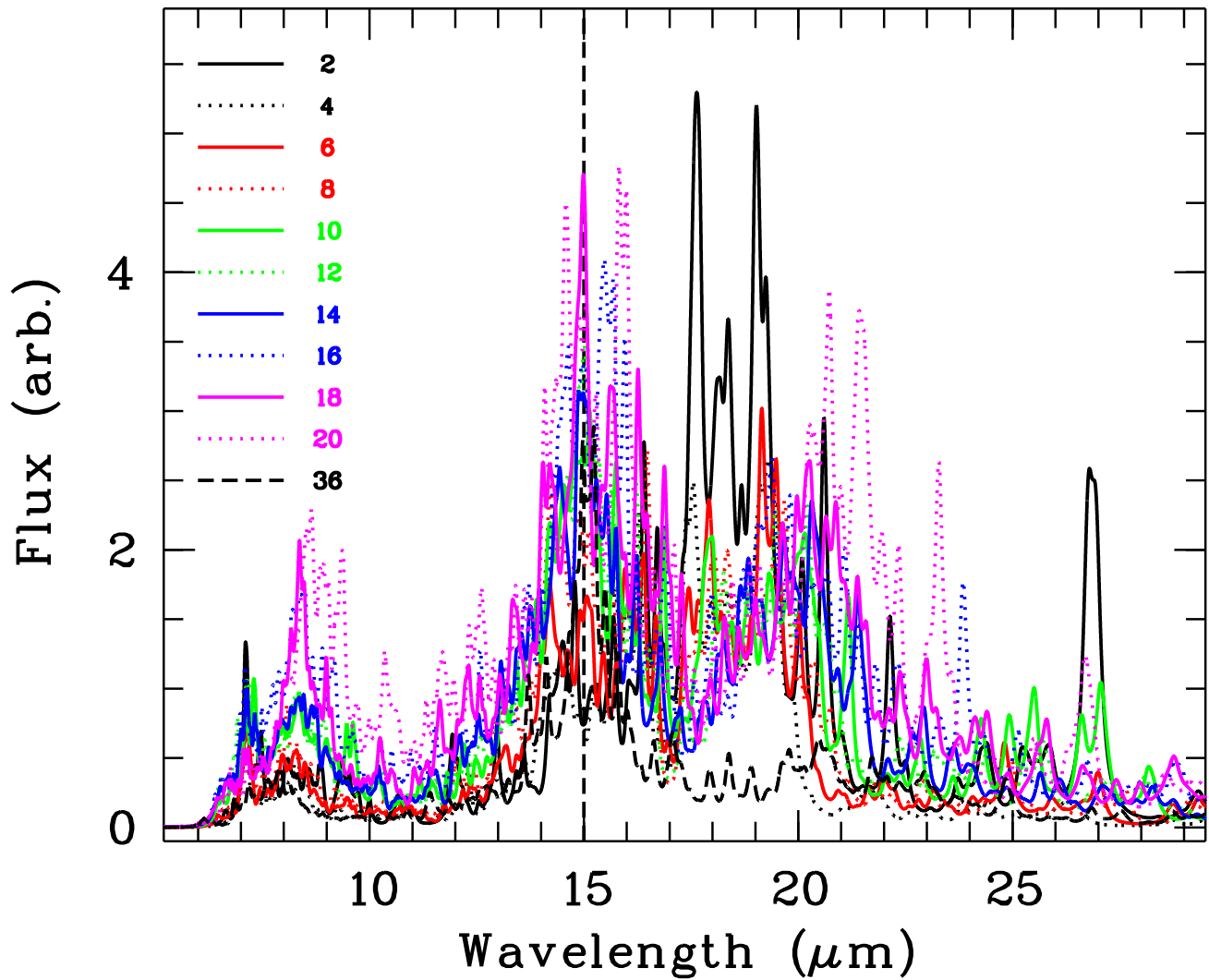}
\caption{%
The theoretical spectra of C$_{60}$H$_m$ computed from a thermal excitation model, where
the temperature is assumed to be 300\,K. The $m$ values are marked on the up-left corner.
A feature around 15\,$\mu$m appears, as marked by the vertical dashed lines
} 
\label{spe}
\end{figure*}

\begin{figure}[tb]
\includegraphics[width=\columnwidth]{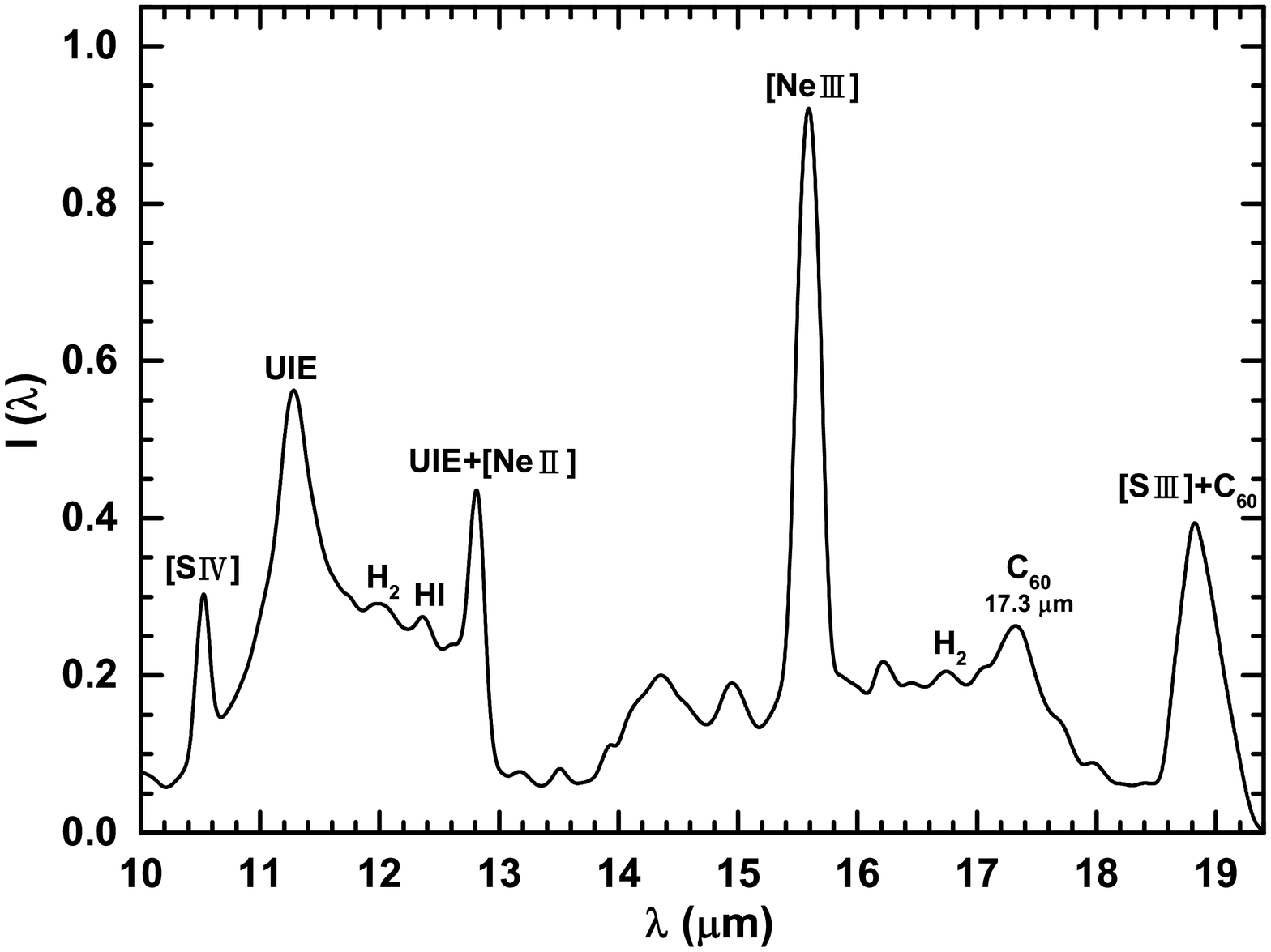}
\caption{%
The stacked {\it Spitzer}/{\it IRS} spectrum of all C$_{60}$ sources in the 10--19\,$\mu$m range
} 
\label{15um}
\end{figure}

Theoretical computations show that the C-H stretching mode is greatly suppressed in cations or
slightly hydrogenated fullerenes \citep{pt95,zhang17}. In such cases, the feature around 
3.4\,$\mu$m is not an ideal proxy to search for fulleranes. Based on quantum chemistry methods
and a thermal excitation model, \citet{zhang17} calculated the mid-infrared
spectra of a number of C$_{60}$H$_m$ isomers with $m=2$--36. 
For each C$_{60}$H$_m$ with a given $m$ value, the coadded spectra of five isomers are presented in 
Figure~\ref{spe}. It appears that two features approximately peaking at 8.5\,$\mu$m and 15\,$\mu$m grow with
hydrogenation. The emission bands around 8.5\,$\mu$m have been commonly discovered in astronomical 
objects, and cannot be uniquely attributed to fulleranes as they can be carried by other aromatic 
hydrocarbon materials. Arising from  the coupling of carbon skeleton modes
and C-H bending motion, the 15\,$\mu$m feature might be able to trace the existence of
fulleranes. However, a thorough search of this feature in C$_{60}$-containing objects
did not yield any solid detection although two of them seem to marginally reveal a peak at 
15\,$\mu$m \citep{zhang17}. Figure~\ref{15um} presents the 
stacked spectrum of all C$_{60}$-containing objects. Although an extremely
faint peak can be seen at 15\,$\mu$m, we cannot firmly conclude that it is
a real feature, rather than noise.
This seems to disfavor the coexistence of 
fulleranes (at least those with high H content) and fullerenes.
However, one cannot completely rule out the possibility that fulleranes exist in condensed phase so 
that the 15\,$\mu$m band is strongly suppressed. Indeed, the laboratory spectrum of C$_{60}$H$_{18}$ 
does not reveal the band at 15\,$\mu$m as strong as that at 8.5\,$\mu$m \citep{igl12}. 

Strikingly, a weak band at 15.8\,$\mu$m was found to be associated with
the 21\,$\mu$m feature \citep{hr08,zhang11,sloan}, which is compatible with the hypothesis of
fulleranes as the carrier of the two features; no 21\,$\mu$m sources have been detected
in C$_{60}$ though. If the  21\,$\mu$m feature is carried by moderately hydrogenated C$_{60}$, this might
suggest that most of the C$_{60}$ molecules have been hydrogenated in the favorable environments of proto-PNe.

With a high proton affinity, C$_{60}$ is readily protonated into C$_{60}$H$^+$ in the PN stage.
\citet{pa19} presented the experimental infrared spectrum of C$_{60}$H$^+$, which exhibits
strong bands in the spectral range of 6.4--8.7\,$\mu$m, mainly corresponding to C-C stretching vibrations.
The authors found a close correspondence
between the experimental spectrum of C$_{60}$H$^+$ and the observed spectra of 
C$_{60}$ PNe in the 6--9\,$\mu$m range.
The C-H stretching band at 3.4\,$\mu$m is not seen in the experimental spectrum, which
is consistent with the non-detection of this feature in C$_{60}$ PNe.

There are other considerations favoring the existence of  C$_{60}$H$_m$. C$_{60}$ was detected in
PNe through four mid-infrared bands, but their relative intensity ratios disagree with the model
predictions of  fluorescent or thermal emission \citep{ber12,zhang13,bri16}. With slight hydrogenation,
the C$_{60}$ cage is distorted and emits the four mid-infrared bands with different intrinsic strengths,
thus providing a natural solution for the discrepancy between the observed and predicted intensity ratios 
\citep{zhang17}.  Indeed, \citet{dw11} proposed that the  C$_{60}$ infrared emission might be 
excited by the energy released via the exothermic reaction of ${\rm H+H \rightarrow H_2}$
on the C$_{60}$  surface. If this is the case, fulleranes should coexist with fullerenes.

It is hard to target specific astronomical sources of searching for fulleranes because
the chemical processes of astronomical fulleranes are rather unclear.
Several formation routes of C$_{60}$ have been proposed including: a)
the formation in H-poor environments \citep{gs92}, b) high-temperature formation
in C-rich environments \citep{jag09}, c) photochemical processing of
hydrogenated amorphous carbon grains \citep{gar10,gar11b,mj12}, d) photochemical
processing of large PAHs \citep{bt12}, and e) shock-induced processing of
SiC grains \citep{ber19}. In these scenarios,  C$_{60}$H$_m$ is not expected to form
together with C$_{60}$; instead, the carbon cage should form at first, and then be hydrogenated  when the objects evolve until
the favorable conditions are achieved.  
When hydrogenation degree increases, the four mid-infrared bands of the C$_{60}$ cage may become invisible. 
Therefore, future searches of fulleranes should also consider the sources that are not detected in C$_{60}$.

\section{Summary}

It was shown that in literature, fulleranes have been proposed to explain almost all
the unidentified features in astronomical spectra, including the extinction bump at 217.5\,nm,
DIBs, ERE, UIE, and AME. Early studies of astronomical fulleranes
did not gain adequate attention because they were largely based on some plausible
conjectures. Recently, the evident discoveries of fullerenes in various astronomical environments 
boost the interests of searching for fulleranes.  It is instructive to review these early attempts
that can provide important guideline and foster future research in this area. 
With remarkable advancements in observational, computational, and experimental techniques,
we are now in a position to revisit the fullerane hypothesis. 
The identification of fulleranes in space has been severely impeded by 
the vast number of fullerane isomers and the absence of laboratory spectra. 
It is worthwhile to construct a database of theoretical and laboratory spectra of various fullerane isomers.
Although previous search efforts are not conclusive, we cannot rule out the existence of 
fulleranes in astronomical environments.  With high infrared sensitivity and high angular resolution,
the upcoming {\it JWST} will provide a unique capability to investigate astronomical fulleranes.

\acknowledgments

We would like to thank an anonymous referee for a very helpful review.
YZ is grateful to the National Science Foundation of China (NSFC, Grant No.  11973099) for the
financial support of this work. CHH acknowledges the Science and Technology Development Fund of 
Macau Special Adminastrative Region for support through grant 0007/2019/A.

\end{document}